\documentclass[twocolumn,showpacs,preprintnumbers,amsmath,amssymb]{revtex4}

\usepackage{graphicx}
\usepackage{dcolumn}
\usepackage{bm}

\begin{document}

\preprint{APS/123-QED}

\title{Stochastic master equation for a probed system in a cavity}

\author{Anne E. B. Nielsen and Klaus M{\o}lmer}
\affiliation{Lundbeck Foundation Theoretical Center for Quantum
System Research, Department of Physics and Astronomy, University of
Aarhus, DK-8000 \AA rhus C, Denmark}

\date{\today}

\begin{abstract}
We present a detailed derivation of the stochastic master equation
determining the time evolution of the state of a general quantum
system, which is placed inside a cavity and subjected to indirect
measurements by monitoring the state of electromagnetic radiation
transmitted through the cavity. The derivation is based on the
physics involved and the final result is stated in terms of the
physical parameters of the setup. To illustrate the predictions
contained in the equation, we solve it analytically for a specific
system, and we demonstrate quantum jumps and freezing of the
internal coherent dynamics of the system as a result of continued
measurements.
\end{abstract}


\maketitle

\section{Introduction}

The time evolution of the state of an isolated quantum mechanical
system is governed by the Schr\"odinger equation. If the system is
subjected to a measurement, the state of the system collapses onto
an eigenstate of the measured observable, and the measurement
outcome is the corresponding eigenvalue. Instead of performing a
measurement directly on the system, one may also allow the system to
interact with an auxiliary system, which is then subjected to a
measurement. This measurement procedure reveals only partial
information on the state of the original system and is very useful
in protocols to prepare a system in a certain quantum mechanical
state and to achieve quantum nondemolition measurements of an
observable. The system could for instance be clouds of atoms or few
atoms that are probed with a beam of light
\cite{kuzmich,duan,geremia,sorensen} or an electromagnetic field
mode in a cavity probed with a beam of atoms \cite{haroche}.

When a system interacts with the surroundings, it is not practical
to keep track of the complete state of both system and surroundings,
and instead the degrees of freedom of the surroundings are traced
out. In this case the state of the system is no longer pure and must
be described by a density operator, whose time evolution is
determined by the master equation. For a system subjected to
indirect measurements, the time evolution depends on the actual
outcomes of the measurements, and since these are probabilistic in
nature, they are handled mathematically by introducing stochastic
variables into the master equation. This allows one to calculate
both the state of the system conditioned on a given set of
measurement outcomes (a given realization of the stochastic
variables) and the probability to observe that particular sequence
of measurement results \cite{jacobs}.

The stochastic master equation is often derived in very general
settings, and it has been shown that all master equations must be on
the so-called Lindblad form \cite{lindblad} in order to preserve
complete positivity of the density operator. The generality is,
however, achieved at the cost of introduction of abstract
measurement operators and measurement strengths. In contrast, in the
present paper we assume a concrete measurement scheme and derive the
stochastic master equation directly from known physical
interactions. The system under consideration is placed inside a
cavity and probed with electromagnetic radiation by sending photons
into the cavity from one side and performing homodyne measurements
on the field leaving the cavity on the opposite side. The purpose of
the cavity is to reflect the probe light several times before it is
detected whereby the effective interaction strength between the
light field and the system is increased as we shall demonstrate
below.

Probing of atomic systems by their effect on the transmission
properties of an optical cavity have a long history in quantum
optics. Early studies focused on the field-atom dynamics, leading,
for instance, to the normal-mode splitting of the transmission
resonance \cite{thompson} and on photon statistics \cite{rempe},
while more recent work has shown the possibility to observe the
spatial motion of individual atoms trapped inside the cavity by the
probing beam itself \cite{pinkse,horak}. See also the work on
optically transported or guided atoms \cite{sauer,fortier,hinds}.
More recently, optical cavities have been introduced in experiments
to probe Bose-Einstein condensates \cite{ottl,brennecke,reichel}.

A stochastic master equation for a setup involving a cavity and a
homodyne detector has been derived using a rather different approach
in Ref.\ \cite{wiseman}. In that paper, however, the aim was not to
use an auxiliary system to perform indirect measurements but to
determine the time evolution of the field in a cavity, when the
light leaking out of the cavity is subjected to measurements, and
thus the probing light and the probed system, which are crucial
ingredients in the present paper, were not included in the analysis.

The paper is structured as follows. In Sec.\ \ref{II} we describe
the physical setup under consideration and discuss the time
evolution of the density matrix. If the dynamics is slow compared to
the round trip time of light in the cavity, we can use a continuous
description and derive the stochastic master equation in Sec.\
\ref{III}. In Sec.\ \ref{IV} we comment further on the derived
equation, and in Sec.\ \ref{V} we apply it to specific systems to
illustrate explicitly how the state of the system is gradually
collapsed by the measurements and to investigate how the relative
strength of the probing and the coherent dynamics of the system
influences the time evolution. Sec.\ \ref{VI} concludes the paper.

\section{Model of the probing procedure}\label{II}

\begin{figure}
\begin{center}
\includegraphics*[viewport=3 3 336 204,width=\columnwidth]{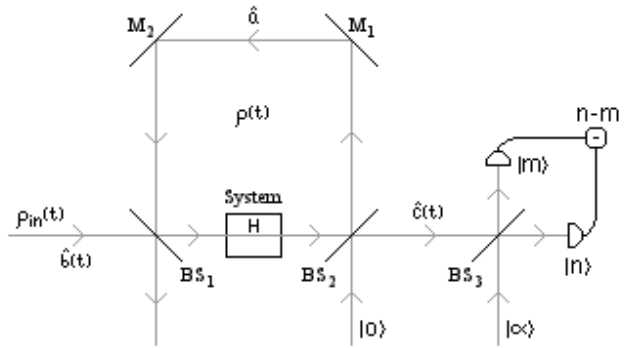}
\end{center}
\caption{Model of the setup. The cavity consists of the two mirrors
M$_1$ and M$_2$ and the two beam splitters BS$_1$ and BS$_2$. The
probe light $\rho_{\textrm{in}}(t)$ enters the cavity through
BS$_1$. The cavity field interacts with the system and leaks out of
the cavity through BS$_1$ and BS$_2$ ($|0\rangle$ is a vacuum
state). The light transmitted through the cavity is subjected to
balanced homodyne detection. BS$_3$ is a 50:50 beam splitter,
$|\alpha\rangle$ is a strong local oscillator, and the two detectors
register $n$ and $m$ photons, respectively. The measurement readout
is $k=n-m$.}\label{setup}
\end{figure}

\begin{figure}
\begin{center}
\includegraphics*[viewport=10 10 338 132,width=0.90\columnwidth]{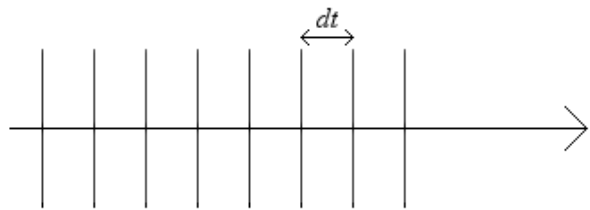}
\end{center}
\caption{The light beams are divided into temporal modes of
(infinitesimal) duration $dt$.}\label{lightmodes}
\end{figure}

The probing procedure applied is shown schematically in Fig.\
\ref{setup}. The system is placed inside a four sided ring cavity.
The probe light enters the cavity from the left, it interacts with
the system, and eventually it leaks out of the cavity, where the
transmitted light is subjected to a balanced homodyne measurement.
In order to describe the quantum state of the light field we divide
all the light beams into small segments of duration $dt$ as
illustrated in Fig.\ \ref{lightmodes} and treat each segment as a
single mode. This is valid provided $dt$ is chosen sufficiently
small. The state of the light and the system is then specified as
the collective state of all the light modes and the system. To avoid
keeping track of an enormous number of modes, we assume that the
probe light is in a coherent state before it enters into the cavity.
The modes of the probe beam that have not yet reached the cavity are
then in a product state, and their time evolution is independent of
the dynamics inside the cavity and the outcome of the homodyne
measurements. These modes can consequently be excluded from the
analysis until the time, when they arrive at the cavity. The vacuum
field incident on the beam splitter BS$_2$ in Fig.\ \ref{setup} and
the coherent state local oscillator field are also product states,
and again it is sufficient to consider the modes that are currently
entering. The modes that are detected in the homodyne detector are
traced out after the measurements, and the unobserved modes that
leave the cavity at BS$_1$ are also traced out in each time step.
The number of active modes is then constant in time, and in the
following we denote the density operator of the light modes inside
the cavity, the light modes between the cavity and the detectors,
and the system at time $t$ by $\rho(t)$. We note that the analysis
can also be carried out for a squeezed input field by including an
optical parametric oscillator in front of the cavity in Fig.\
\ref{setup}. For a suitable theoretical description of the optical
parametric oscillator see Ref.\ \cite{nielsen4}.

To determine the time evolution of $\rho(t)$ we express $\rho(t+dt)$
in terms of $\rho(t)$. The relevant modes are the modes included in
$\rho(t)$ and the mode of the probe field, the vacuum state mode,
and the local oscillator mode that reach beam splitters BS$_1$,
BS$_2$, and BS$_3$, respectively, in the time interval between $t$
and $t+dt$, and the state of these modes is $\rho(t)\otimes
\rho_{\textrm{in}}(t)\otimes|0\rangle\langle0|\otimes|\alpha\rangle
\langle\alpha|$. In the time between $t$ and $t+dt$ several
interactions take place. The state of the two modes that hit the
beam splitter BS$_1$ are mixed, and this process is described by the
unitary operator $U_1$. Similar transformations, described by the
operators $U_2$ and $U_3$, occur at beam splitters BS$_2$ and
BS$_3$. The system itself undergoes evolution during $dt$ and the
light mode that passes the system at time $t$ interacts with it. The
corresponding infinitesimal time evolution operator is
$U_H=1-iHdt/\hbar$, where $H=H_{\textrm{sys}}+H_I$ is the sum of the
system Hamiltonian $H_{\textrm{sys}}$ and the Hamiltonian $H_I$ for
the interaction between the system and the light field. We note that
the system could be subjected to manipulations that depend on the
state of the system and light field and such feedback terms would be
included in $H_{\textrm{sys}}$. In the following we disregard decay
of the system, since it does not add any interesting points to the
analysis, and it is easily included by adding a term to the final
equation (see for instance \cite{jacobs} for a derivation of the
relevant term for a system consisting of a single two-level atom).
If the system is an extended object that interacts strongly and, for
example, depletes the light field, it would be relevant to slice the
system into small pieces and consider interactions between the
system and several light modes in each time step, but we ignore such
complications here. We shall also assume that the mirrors M$_1$ and
M$_2$ induce phase shifts of $\pi/2$, which are taken into account
through the operators $U_{M_1}$ and $U_{M_2}$. Finally, the two
modes that hit the detectors in the time interval $dt$ are projected
on photon number states $|n\rangle$ and $|m\rangle$, respectively.
Even though the detectors are able to resolve the exact number of
photons microscopically, the only macroscopically available
measurement readout is the difference $k=n-m$ between the number of
detected photons in the two detectors. We thus have to average over
all possible values of $n$ and $m$ that lead to the observed value
of $k$. Putting all the transformations together, we obtain the
density operator at time $t+dt$ conditioned on the measurement of a
difference of $k$ photons in the interval from $t$ to $t+dt$
\begin{multline}\label{timeevolution}
\rho(t+dt)=\frac{1}{P_k}\textrm{Tr}_1\Big(\sum_m\langle m+k|\langle m|U_3U_1U_{M_2}\\
U_{M_1}U_2U_H\rho(t)\otimes\rho_{\textrm{in}}(t)\otimes|0\rangle\langle0|
\otimes|\alpha\rangle\langle\alpha|\\
U^\dag_HU^\dag_2U^\dag_{M_1}U^\dag_{M_2}U^\dag_1
U^\dag_3|m\rangle|m+k\rangle\Big),
\end{multline}
where $P_k$ is the probability, determined from the normalization of
$\rho(t+dt)$, to obtain the measurement outcome $k$, and Tr$_1$
denotes the trace over the unobserved mode leaving the cavity at
beam splitter BS$_1$.

The time evolution of $\rho(t)$ is completely specified by Eq.\
\eqref{timeevolution}, and in principle one can start with a given
initial state and iterate \eqref{timeevolution} on a computer. This
task does, however, soon become unwieldy due to the large number of
modes that are involved. In the next section we show how Eq.\
\eqref{timeevolution} can be simplified considerably and rewritten
as a stochastic differential equation by invoking a few assumptions.

\section{Derivation of the stochastic master equation}\label{III}

The key assumption in the following derivation is that all changes
of the state of the light field takes place on a time scale that is
large compared to the round trip time $\tau$ of light in the cavity,
which is valid if the cavity is sufficiently small. In this case the
temporal width of the light modes can be chosen as large as $\tau$,
and, in particular, the field inside the cavity can be treated as a
single mode. The assumption requires that the coupling between the
system and the light field is not too large, since, for instance,
the fraction of the light absorbed or emitted during a single round
trip must be small. The finesse of the cavity must also be high and
the total number of photons in a segment of duration $\tau$ of the
input beam must be small in order to avoid that a significant number
of photons leak out of or enter into the cavity within a time
$\tau$. Finally, variations in, for instance, the input field or the
interactions between the system and the light field that take place
on a time scale small compared to $\tau$ are not allowed, since they
will be smeared out. Actually, the input field typically has such
fast variations because the size of the cavity is large compared to
the wavelength of the input light, but, provided the frequency of
the radiation is not too far from a cavity resonance, this problem
is easily circumvented by moving into a frame rotating with the
relevant resonance frequency of the cavity, and in the following all
frequencies are to be measured relative to the cavity resonance.
With these requirements we can treat $\tau$ as an infinitesimal
quantity, and below we shall denote $\tau$ by $dt$. For simplicity,
we shall also assume that the homodyne detector is placed within a
distance $c\tau$ from BS$_2$, where $c$ is the speed of light, since
in that case $\rho(t)$ includes only the state of the system and a
single mode of the light field (the cavity mode).

When $\tau$ is infinitesimal, the transmission $t_1^2$ ($t_2^2$) of
beam splitter BS$_1$ (BS$_2$) must also be infinitesimal, and we
define $\kappa_i$ by
\begin{equation}\label{kappa}
t_i^2=\kappa_idt,\hspace{1cm}i=1,2.
\end{equation}
This allows us to rewrite the unitary operator
\begin{equation}\label{U1}
U_1=e^{i\frac{\pi}{2}\hat{a}^\dag\hat{a}}e^{i\frac{\pi}{2}\hat{b}^\dag\hat{b}}
e^{-i\tan^{-1}\left(\frac{t_1}{r_1}\right)
\left(\hat{a}^\dag\hat{b}+\hat{a}\hat{b}^\dag\right)},
\end{equation}
representing BS$_1$, as
\begin{multline}\label{U1app}
U_1=e^{i\frac{\pi}{2}\hat{a}^\dag\hat{a}}e^{i\frac{\pi}{2}\hat{b}^\dag\hat{b}}\\
\left(1-it_1(\hat{a}^\dag\hat{b}+\hat{a}\hat{b}^\dag)
-\frac{1}{2}t_1^2(\hat{a}^\dag\hat{b}+\hat{a}\hat{b}^\dag)^2\right)
\end{multline}
to first order in $dt$, where $r_1^2$ is the reflectivity of BS$_1$,
$\hat{a}$ is the field annihilation operator of the cavity field,
and $\hat{b}$ is the field annihilation operator of the mode of
width $dt$ of the input field that arrives at the beam splitter at
time $t$. Note that \eqref{U1} has been chosen such that after the
interaction, the $\hat{a}$-mode is still the cavity mode, while the
$\hat{b}$-mode is the mode leaving the cavity. Similarly, we have
\begin{multline}\label{U2app}
U_2=e^{i\frac{\pi}{2}\hat{a}^\dag\hat{a}}e^{i\frac{\pi}{2}\hat{c}^\dag\hat{c}}\\
\left(1-it_2(\hat{a}^\dag\hat{c}+\hat{a}\hat{c}^\dag)
-\frac{1}{2}t_2^2(\hat{a}^\dag\hat{c}+\hat{a}\hat{c}^\dag)^2\right),
\end{multline}
where $\hat{c}$ is the field annihilation operator of the mode
leaving the cavity at BS$_2$. The operators $U_{M_1}$ and $U_{M_2}$
both act on the cavity mode and thus
\begin{equation}\label{UM}
U_{M_1}=U_{M_2}=e^{i\frac{\pi}{2}\hat{a}^\dag\hat{a}}.
\end{equation}
Since the number of photons in a segment of duration $\tau$ of the
input beam is assumed to be much smaller than one, we can write
\begin{multline}\label{rhoin}
\rho_{\textrm{in}}(t)=c_{\textrm{in},00}|0\rangle\langle0|
+c_{\textrm{in},10}|1\rangle\langle0|
+c_{\textrm{in},01}|0\rangle\langle1|\\
+c_{\textrm{in},11}|1\rangle\langle1|
+c_{\textrm{in},20}|2\rangle\langle0|
+c_{\textrm{in},02}|0\rangle\langle2|
\end{multline}
to first order in $dt$, where
$c_{\textrm{in},00}=1-c_{\textrm{in},11}$ is of order unity,
$c_{\textrm{in},10}=c_{\textrm{in},01}^*$ is of order $\sqrt{dt}$,
and $c_{\textrm{in},11}$ and
$c_{\textrm{in},20}=c_{\textrm{in},02}^*$ are of order $dt$.

Putting the relation $U_H=1-iHdt/\hbar$ and Eqs.\ \eqref{U1app},
\eqref{U2app}, \eqref{UM}, and \eqref{rhoin} into Eq.\
\eqref{timeevolution}, we find
\begin{multline}\label{rhotdt}
\rho(t+dt)=\frac{1}{P_k}\bigg(u_{00}\rho(t)
-u_{00}\frac{i}{\hbar}[H,\rho(t)]dt+u_{00}t_1\\
\left([\hat{a}^\dag,\rho(t)]\mathrm{Tr}\left(\hat{b}\rho_{in}(t)\right)
-[\hat{a},\rho(t)]\mathrm{Tr}\left(\rho_{in}(t)\hat{b}^\dag\right)\right)\\
+u_{10}t_2\hat{a}\rho(t)+u_{01}t_2\rho(t)\hat{a}^\dag\\
+\frac{1}{2}u_{00}t_1^2\left(-\hat{a}^\dag\hat{a}\rho(t)
-\rho(t)\hat{a}^\dag\hat{a}+2\hat{a}\rho(t)\hat{a}^\dag\right)\\
+\frac{1}{2}t_2^2\big(-u_{00}\hat{a}^\dag\hat{a}\rho(t)
-u_{00}\rho(t)\hat{a}^\dag\hat{a}
+2u_{11}\hat{a}\rho(t)\hat{a}^\dag\\
+\sqrt{2}u_{20}\hat{a}^2\rho(t)
+\sqrt{2}u_{02}\rho(t)(\hat{a}^\dag)^2\big)\bigg),
\end{multline}
where
\begin{equation}\label{u}
u_{pq}\equiv\sum_m\langle m+k|\langle
m|U_3|\alpha\rangle|p\rangle\langle
q|\langle\alpha|U^\dag_3|m\rangle|m+k\rangle
\end{equation}
and $|p\rangle$ and $|q\rangle$ are photon number states. For a
strong local oscillator ($|\alpha|^2\gg1$) we can apply the
approximation
\begin{equation}
\frac{1}{n!}\mu^ne^{-\mu}\approx
\frac{1}{\sqrt{2\pi\mu}}e^{-\frac{(n-\mu)^2}{2\mu}},
\hspace{7mm}\mu\equiv\frac{|\alpha|^2}{2}
\end{equation}
and turn the sum in Eq.\ \eqref{u} into an integral, which leads to
($\alpha=|\alpha|e^{i\phi}$)
\begin{eqnarray}
u_{00}&=&
\frac{1}{2\sqrt{\pi\mu}}\exp\left(-\frac{k^2}{4\mu}\right),\label{u00}\\
u_{10}&=& \frac{-ike^{-i\phi}}{\sqrt{2\mu}}\frac{1}{2\sqrt{\pi\mu}}
\exp\left(-\frac{k^2}{4\mu}\right),\label{u10}\\
u_{20}&=&-\frac{(k^2-2\mu)e^{-2i\phi}}{2\sqrt{2}\mu}
\frac{1}{2\sqrt{\pi\mu}}\exp\left(-\frac{k^2}{4\mu}\right),\label{u20}\\
u_{11}&=&
\frac{k^2}{2\mu}\frac{1}{2\sqrt{\pi\mu}}\exp\left(-\frac{k^2}{4\mu}\right)\label{u11}.
\end{eqnarray}
From the normalization of Eq.\ \eqref{rhotdt} we find the
probability to obtain the measurement outcome $k$
\begin{multline}\label{Pk}
P_k=\frac{1}{2\sqrt{\pi\mu}}\exp\left(-\frac{k^2}{4\mu}\right)
\bigg(1-\frac{ike^{-i\phi}}{\sqrt{2\mu}}t_2\textrm{Tr}\left(\hat{a}\rho(t)\right)\\
+\frac{ike^{i\phi}}{\sqrt{2\mu}}t_2\textrm{Tr}\left(\rho(t)\hat{a}^\dag\right)
+\frac{1}{2}t_2^2\left(\frac{k^2}{2\mu}-1\right)\Big(
2\textrm{Tr}\left(\hat{a}\rho(t)\hat{a}^\dag\right)\\
-e^{-2i\phi}\textrm{Tr}\left(\hat{a}^2\rho(t)\right)
-e^{2i\phi}\textrm{Tr}\left(\rho(t)(\hat{a}^\dag)^2\right)\Big)\bigg).
\end{multline}
Comparing this to
\begin{multline}
\frac{1}{2\sqrt{\pi(\mu+\epsilon)}}\exp\left(-\frac{(k-\delta)^2}{4(\mu+\epsilon)}\right)
\approx\\
\frac{1}{2\sqrt{\pi\mu}}\exp\left(-\frac{k^2}{4\mu}\right)
\left(1+\frac{k\delta}{2\mu}
+\left(\frac{k^2}{2\mu}-1\right)\frac{\epsilon}{2\mu}\right)
\end{multline}
for $\delta\ll k$ and $\epsilon\ll\mu$, it is apparent that $P_k$,
to first order in $dt$, is a Gaussian distribution. $\delta$ is of
order $\sqrt{dt}$, and $\epsilon$ is of order $dt$, so to order
$\sqrt{dt}$ we can replace $k/\sqrt{2\mu}$ by
\begin{equation}\label{k}
\frac{\hat{k}}{\sqrt{2\mu}}=-ie^{-i\phi}t_2\textrm{Tr}\left(\hat{a}\rho(t)\right)
+ie^{i\phi}t_2\textrm{Tr}\left(\rho(t)\hat{a}^\dag\right)
+\frac{d\hat{W}}{\sqrt{dt}},
\end{equation}
where $d\hat{W}$ is a stochastic variable that has a Gaussian
probability density distribution with zero mean value and variance
$dt$, and we have put a hat on $k$ to emphasize that $\hat{k}$ is
now to be regarded as a stochastic variable that assumes the value
$k$ with probability $P_k$. From the Ito calculus rule
$d\hat{W}^2=dt$ (see \cite{jacobs}) it furthermore follows that
\begin{equation}\label{k2}
\hat{k}^2=2\mu
\end{equation}
to zeroth order in $dt$. Inserting Eqs.\ \eqref{kappa}, \eqref{u00},
\eqref{u10}, \eqref{u20}, \eqref{u11}, \eqref{Pk}, \eqref{k},
\eqref{k2} and
\begin{equation}
\textrm{Tr}\left(\hat{b}\rho_{\textrm{in}}(t)\right)
=\textrm{Tr}\left(\rho_{\textrm{in}}(t)\hat{b}^\dag\right)^*
=c_{\textrm{in},10}\equiv\beta\sqrt{dt}
\end{equation}
into Eq.\ \eqref{rhotdt} we finally obtain the stochastic master
equation in the form
\begin{multline}\label{SME}
\rho(t+dt)=\rho(t)-\frac{i}{\hbar}\left[H,\rho(t)\right]dt\\
+\sqrt{\kappa_1}[\hat{a}^\dag,\rho(t)]\beta dt
-\sqrt{\kappa_1}[\hat{a},\rho(t)]\beta^* dt\\
-ie^{-i\phi}\sqrt{\kappa_2}
\left(\hat{a}\rho(t)-\mathrm{Tr}\left(\hat{a}\rho(t)\right)\rho(t)\right)d\hat{W}\\
+ie^{i\phi}\sqrt{\kappa_2}\left(\rho(t)\hat{a}^\dag-
\mathrm{Tr}\left(\rho(t)\hat{a}^\dag\right)\rho(t)\right)d\hat{W}\\
+\frac{1}{2}\left(\kappa_1+\kappa_2\right)\left(-\hat{a}^\dag\hat{a}\rho(t)-
\rho(t)\hat{a}^\dag\hat{a}+2\hat{a}\rho(t)\hat{a}^\dag\right)dt.
\end{multline}
The second term on the right hand side represents the evolution due
to the system Hamiltonian and the interaction between the system and
the radiation, the third and fourth terms arise due to the feeding
of probe light into the cavity at BS$_1$, the fifth and sixth terms
include the knowledge obtained from the homodyne measurements, and
the seventh term describes the decay of the cavity mode due to
transmission through BS$_1$ and BS$_2$. We note that the
approximations leading to Eq.\ \eqref{SME} can be stated more
precisely as $|\alpha|^2\gg1$, $\kappa_1\tau\ll1$,
$\kappa_2\tau\ll1$, $\sqrt{\kappa_1}|\beta|\tau\ll1$, and
$|\langle\psi_1|H|\psi_2\rangle|\tau/\hbar\ll1$, where
$|\psi_1\rangle$ and $|\psi_2\rangle$ represent arbitrary state
kets. It is allowed that $\beta$ and $H$ are time dependent, but the
variation within a time interval of length $\tau$ should be small.

\section{Further remarks}\label{IV}

So far we have assumed a lossless setup, but it is easy to
incorporate effects of losses. To account for losses in the light
field, we only need to replace the perfect mirror M2 by a partially
transmitting beam splitter and include a beam splitter in front of
the ideal homodyne detector with transmissivity $\eta=\eta_D\eta_P$,
where $\eta_D$ is the efficiency of the detector and $1-\eta_P$ is
the propagation loss between the cavity and the detector, and in
this case Eq.\ \eqref{SME} generalizes to
\begin{multline}\label{SMET}
\rho(t+dt)=\rho(t)-\frac{i}{\hbar}\left[H,\rho(t)\right]dt\\
+\sqrt{\kappa_1}\left[\hat{a}^\dag,\rho(t)\right]\beta dt
-\sqrt{\kappa_1}\left[\hat{a},\rho(t)\right]\beta^*dt\\
-ie^{-i\phi}\sqrt{\eta\kappa_2}
\left(\hat{a}\rho(t)-\mathrm{Tr}\left(\hat{a}\rho(t)\right)\rho(t)\right)d\hat{W}\\
+ie^{i\phi}\sqrt{\eta\kappa_2}
\left(\rho(t)\hat{a}^\dag-\mathrm{Tr}
\left(\rho(t)\hat{a}^\dag\right)\rho(t)\right)d\hat{W}\\
+\frac{1}{2}\kappa\left(-\hat{a}^\dag\hat{a}\rho(t)-
\rho(t)\hat{a}^\dag\hat{a}+2\hat{a}\rho(t)\hat{a}^\dag\right)dt,
\end{multline}
where $\kappa\equiv\kappa_1+\kappa_2+\kappa_L$, and $\kappa_L$,
defined in analogy to Eq.\ \eqref{kappa}, is the decay rate due to
intra cavity loss. It is also possible to account for loss and
decoherence of the quantum system in the cavity by introducing
appropriate damping terms in Eq.\ \eqref{SMET}.

If there is no decay of the system and no cavity loss and all light
emerging from inside the cavity is detected with unit efficiency
detectors, the dynamics will preserve the purity of an initially
pure state of the atoms and the cavity field, and the stochastic
master equation can be rewritten as a stochastic Schr\"odinger
equation. This is convenient since it is significantly easier to
propagate a wave function in time than a density operator. The
situation can be achieved with the setup in Fig.\ \ref{setup} by
taking the limit $\kappa_1\rightarrow0$ and
$|\beta|\rightarrow\infty$ while $\sqrt{\kappa_1}|\beta|\tau$ is
kept small compared to unity. Alternatively one could measure both
the light reflected and transmitted from the cavity or replace
$\textrm{BS}_2$ by a perfectly reflecting mirror and subject the
light leaving the cavity at $\textrm{BS}_1$ to homodyne detection.
In the latter case the stochastic master equation is given by Eq.\
\eqref{SME} with $\kappa_2$ and $\kappa_1+\kappa_2$ replaced by
$\kappa_1$ and $\phi$ replaced by $\phi+\pi/2$, and the stochastic
Schr\"odinger equation reads
\begin{multline}\label{SSE}
|\psi(t+dt)\rangle=\Big(1-\frac{i}{\hbar}Hdt+\beta\sqrt{\kappa_1}\hat{a}^\dag
dt-\beta^*\sqrt{\kappa_1}\hat{a}dt\\
-e^{-i\phi}\sqrt{\kappa_1}(\hat{a}-\langle\hat{a}\rangle)dW\\
-\frac{\kappa_1}{2}(\hat{a}^\dag\hat{a}
-2\hat{a}\langle\hat{a}^\dag\rangle
+\langle\hat{a}\rangle\langle\hat{a}^\dag\rangle)dt\Big)|\psi(t)\rangle,
\end{multline}
where $\langle\cdot\rangle$ denotes expectation value.

Equation \eqref{SMET} is a nonlinear equation since
$\mathrm{Tr}\left(\hat{a}\rho(t)\right)$ and
$\mathrm{Tr}\left(\rho(t)\hat{a}^\dag\right)$ depend on the state
$\rho(t)$. In general, $H$ may also depend on $\rho(t)$ (for
instance if a state dependent feedback is applied), but in the
special case of a state independent Hamiltonian, it is possible to
transform Eq.\ \eqref{SMET} into a linear equation by application of
the method presented in Ref.\ \cite{jacobs}. The nonlinear terms in
Eq.\ \eqref{SMET} appear due to the first two terms on the right
hand side of Eq.\ \eqref{k} and due to the normalization of the
state. If we simply remove the factor $1/P_k$ in Eq.\
\eqref{rhotdt}, the norm of $\rho(t+dt)$ is the probability to
obtain the measurement outcome $k$ for the time interval between $t$
and $t+dt$. Furthermore, the right hand side of the master equation
is determined completely by the value $k$ assumed by $\hat{k}$ (or
equivalently by the value $dW$ assumed by $d\hat{W}$) and is thus
not changed if the probability density distribution of $d\hat{W}$ is
changed. Only the probability to obtain the value of $d\hat{W}$ that
leads to a specific value of $\hat{k}$ is changed. We can thus
change the probability density distribution of $d\hat{W}$ provided
we accept that the probability to obtain the state $\rho(t+dt)$ at
time $t+dt$ is the norm of the state and not the probability to
obtain the required value of $d\hat{W}$. We exploit this freedom to
define the new stochastic variable $d\hat{y}$ by
\begin{equation}\label{dy}
\frac{\hat{k}}{\sqrt{2\mu}}=\frac{d\hat{y}}{\sqrt{dt}}
\end{equation}
and assume that the probability density distribution of $d\hat{y}$
is a Gaussian distribution with zero mean value and variance $dt$.
If we omit the factor $1/P_k$ in Eq.\ \eqref{rhotdt} and insert
Eqs.\ \eqref{u00}, \eqref{u10}, \eqref{u20}, \eqref{u11}, and
\eqref{dy}, we obtain a linear equation in $\rho(t)$. The trace of
$\rho(t+dt)$ integrated over all possible realizations of $\hat{k}$
is unity, but since $\hat{k}$ is expressed in terms of $d\hat{y}$,
we would like the trace of $\rho(t+dt)$ integrated over all possible
realizations of $d\hat{y}$ to be unity, and we thus multiply the
right hand side by $\sqrt{2\mu/dt}$. Finally, to ensure that
$\rho(t+dt)$ approaches $\rho(t)$ in the limit, where $dt$
approaches zero, we divide the right hand side by
$\exp\left(-k^2/(4\mu)\right)/\sqrt{2\pi dt}$, which is precisely
the Gaussian probability density $P_{dy}$ for $d\hat{y}$. With these
changes Eq.\ \eqref{SMET} reduces to
\begin{multline}\label{SMEL}
\rho(t+dt)=\rho(t)-\frac{i}{\hbar}\left[H,\rho(t)\right]dt\\
+\sqrt{\kappa_1}\left[\hat{a}^\dag,\rho(t)\right]\beta dt
-\sqrt{\kappa_1}\left[\hat{a},\rho(t)\right]\beta^*dt\\
-i\sqrt{\eta\kappa_2}\left(\hat{a}\rho(t)
e^{-i\phi}-\rho(t)\hat{a}^\dag
e^{i\phi}\right)d\hat{y}\\
+\frac{1}{2}\kappa\left(-\hat{a}^\dag\hat{a}\rho(t)-
\rho(t)\hat{a}^\dag\hat{a}+2\hat{a}\rho(t)\hat{a}^\dag\right)dt.
\end{multline}
Due to the division by $P_{dy}$, the probability density to obtain a
given state $\rho(t+dt)$ at time $t+dt$ is now the product of the
norm of $\rho(t+dt)$ and the probability density $P_{dy}$ to obtain
the required value of $d\hat{y}$. Since the stochastic variables
corresponding to different time steps are independent and since Eq.\
\eqref{SMEL} is linear in $\rho(t)$, the probability density to
obtain the specific state after $N$ time steps is simply the
probability  $\prod_{i=1}^NP_{dy_i}$ to obtain the realizations
$dy_1$, $dy_2$, $\ldots$ , $dy_N$ of the stochastic variables
multiplied by the norm of the state obtained from this realization
summed over all realizations that lead to the desired state. In
particular, if the state after $N$ time steps only depends on the
sum $\hat{y}=\sum_{i=1}^Nd\hat{y}_i$, then the probability density
to obtain this state is
\begin{equation}\label{py}
P=\frac{1}{\sqrt{2\pi Ndt}}
\exp\left(-\frac{y^2}{2Ndt}\right)\textrm{Tr}(\rho(t+Ndt)).
\end{equation}

In the preceding section we have assumed that the transmitted light
is observed by a homodyne detector, but a similar derivation may be
carried out for an avalanche photo diode detector, which projects
the infinitesimal modes of the transmitted light on either vacuum or
a single photon state, and for completeness we state the result
\begin{multline}
\rho(t+dt)=\rho(t)-\frac{i}{\hbar}[H,\rho(t)]dt\\
+\sqrt{\kappa_1}\left[\hat{a}^\dag,\rho(t)\right]\beta dt
-\sqrt{\kappa_1}\left[\hat{a},\rho(t)\right]\beta^*dt\\
-\eta\kappa_2(\hat{a}\rho(t)\hat{a}^\dag-
\textrm{Tr}(\hat{a}\rho(t)\hat{a}^\dag)\rho(t))dt\\
+\left(\frac{\hat{a}\rho(t)\hat{a}^\dag}
{\textrm{Tr}(\hat{a}\rho(t)\hat{a}^\dag)}-\rho(t)\right)d\hat{N}\\
+\frac{1}{2}\kappa(-\hat{a}^\dag\hat{a}\rho(t)
-\rho(t)\hat{a}^\dag\hat{a}+2\hat{a}\rho(t)\hat{a}^\dag)dt.
\end{multline}
$d\hat{N}$ is a stochastic variable, which assumes the value $1$
with probability
$\eta\kappa_2\textrm{Tr}(\hat{a}\rho(t)\hat{a}^\dag)dt$ and the
value $0$ with probability
$1-\eta\kappa_2\textrm{Tr}(\hat{a}\rho(t)\hat{a}^\dag)dt$.

\section{Applications}\label{V}

\subsection{The empty cavity}\label{VA}

We first consider the time evolution of the state of the light field
in the absence of any quantum system in the cavity. This situation
leads to a coherent state cavity field because beam splitters
transform coherent states into coherent states and because homodyne
measurements also preserve the coherent state nature of the cavity
field (see e.g.\ Ref.\ \cite{mm}). Inserting $H=0$ and
$\rho(t)=C(t)|\xi(t)\rangle\langle\xi(t)|$ in Eq.\ \eqref{SMET}, we
obtain
\begin{equation}
\frac{d\xi(t)}{dt}=-\frac{\kappa}{2}\xi(t)+\sqrt{\kappa_1}\beta,
\end{equation}
and if $\beta$ is time independent, which implies that the input
light is on resonance with the cavity, $\xi(t)$ approaches
$\xi=2\sqrt{\kappa_1}\beta/\kappa$ for $t\gg2/\kappa$.

We note that $\xi$ is a factor of
$2\sqrt{\kappa_1}/(\sqrt{\tau}\kappa)$ larger than the coherent
state amplitude $\sqrt{\tau}\beta$ of a segment of the input field
of length $c\tau$, and, in the limit of very weak coupling between
the light field and a system, the presence of the cavity enhances
the Rabi frequency of transitions between different states of the
system by the same factor. For $\kappa_3=0$ and $\kappa_1=\kappa_2$
the factor reduces to $1/t_2$, which equals the square root of twice
the average number of round trips of a photon in the cavity in
absence of the system and in absence of the input field. The factor
of two appears due to destructive interference between the input
field and the cavity field for
$\xi(t)=2\sqrt{\kappa_1}\beta/\kappa$, which ensures that no light
is lost at beam splitter BS$_1$ in the presence of the input light.

\subsection{Analytical solution for a simple system}

As a nontrivial application of the stochastic master equation we
analyze the setup proposed in Ref.\ \cite{sorensen} to generate
Dicke states. Here the system consists of $N$ noninteracting
identical atoms each with two ground state levels $|f\rangle$ and
$|g\rangle$. The cavity field couples the level $|f\rangle$ to an
exited level $|e\rangle$, and below it is assumed that the radiation
is sufficiently off-resonant to avoid population of the exited
level. In that case the interaction merely shifts the phase of the
light field by an amount that is proportional to the number of atoms
in the state $|f\rangle$, and since the homodyne detector is
sensitive to the phase shift, the state of the monitored system
slowly approaches an eigenstate of the operator $\hat{n}$, which
counts the number of atoms in the state $|f\rangle$. For atoms all
coupling with equal strengths to the light field, the relevant
Hamiltonian reads
\begin{equation}\label{H}
H=\hbar g \hat{a}^\dag\hat{a}\hat{n},
\end{equation}
where $g$ represents the strength of the coupling. We assume that
the atoms are initially prepared in a state, which is symmetric
under exchange of any two atoms, and we denote the symmetric state
with $n$ atoms in the state $|f\rangle$ by $|n\rangle$. Since the
Hamiltonian \eqref{H} preserves the symmetry under exchange of any
two atoms, this allows us to use the restricted basis consisting of
the states $|n\rangle$ with $n=0,1,\ldots,N$.

To simplify the stochastic master equation, we write the density
operator of the system and the cavity mode as
\begin{equation}
\rho(t)=\sum_{n=0}^N\sum_{m=0}^N\rho_{nm}\otimes|n\rangle\langle m|,
\end{equation}
where $|n\rangle$ and $|m\rangle$ refer to the state of the system
and $\rho_{nm}$ are time dependent cavity mode operators. Inserting
this into Eq.\ \eqref{SMEL}, we obtain the $(N+1)^2$ independent
linear equations
\begin{multline}\label{rhonm}
\frac{d\rho_{nm}}{dt}=-ig(n\hat{a}^\dag\hat{a}\rho_{nm}
-m\rho_{nm}\hat{a}^\dag\hat{a})\\
+\sqrt{\kappa_1}\left[\hat{a}^\dag,\rho_{nm}\right]\beta
-\sqrt{\kappa_1}\left[\hat{a},\rho_{nm}\right]\beta^*\\
-i\sqrt{\eta\kappa_2}\left(\hat{a}\rho_{nm}e^{-i\phi}
-\rho_{nm}\hat{a}^\dag e^{i\phi}\right)\frac{d\hat{y}}{dt}\\
+\frac{1}{2}\kappa\left(-\hat{a}^\dag\hat{a}\rho_{nm}
-\rho_{nm}\hat{a}^\dag\hat{a}+2\hat{a}\rho_{nm}\hat{a}^\dag\right).
\end{multline}
To determine $\langle n|\rho_{\textrm{sys}}|n\rangle$, where
$\rho_{\textrm{sys}}$ is the normalized density operator for the
system, it is sufficient to solve Eq.\ \eqref{rhonm} for $n=m$. If
the initial state of the system is $|n\rangle\langle n|$, we can
replace the operator $\hat{n}$ in the Hamiltonian \eqref{H} by the
number $n$, whereby the system is effectively reduced to a phase
shifter, and it follows that the cavity field is in a coherent
state. Equation \eqref{rhonm} for $n=m$ is, however, not
mathematically different in the general case, and it is thus solved
by
\begin{equation}\label{rhonn}
\rho_{nn}=C_n(t)|\xi_n(t)\rangle\langle\xi_n(t)|.
\end{equation}
Inserting Eq.\ \eqref{rhonn} into Eq.\ \eqref{rhonm}, we find
\begin{equation}
\frac{d\xi_n(t)}{dt}=-\left(\frac{\kappa}{2}+ign\right)\xi_n(t)
+\sqrt{\kappa_1}\beta
\end{equation}
and
\begin{equation}
dC_n(t)=\sqrt{\eta\kappa_2}
\left(-ie^{-i\phi}\xi_n(t)+ie^{i\phi}\xi_n^*(t)\right)C_n(t)d\hat{y},
\end{equation}
and thus, for a time independent $\beta$,
\begin{equation}
\xi_n(t)=\left(\xi_n(0)-\frac{2\sqrt{\kappa_1}\beta}{\kappa+2ign}\right)
e^{-(\kappa/2+ign)t}+\frac{2\sqrt{\kappa_1}\beta}{\kappa+2ign}
\end{equation}
and
\begin{multline}\label{cn}
C_n(t)=C_n(0)\exp\Bigg(-\int_0^tr_n(t')\frac{d\hat{y}(t')}{dt'}dt'\\
-\frac{1}{2}\int_0^tr_n(t')^2dt'\Bigg),
\end{multline}
where $r_n(t)\equiv\sqrt{\eta\kappa_2}
\left(ie^{-i\phi}\xi_n(t)-ie^{i\phi}\xi_n^*(t)\right)$.

The general solution simplifies considerably if the detector is
turned off ($\eta=0$) until $\xi_n(t)$ has obtained its steady state
value
\begin{equation}\label{eq}
\xi_n=\frac{2\sqrt{\kappa_1}\beta}{\kappa}\frac{1-2ign/\kappa}
{1+4g^2n^2/\kappa^2},
\end{equation}
since in that case $r_n(t')=r_n$ can be moved outside the integrals,
and $C_n(t)$ depends only on the sum $\hat{y}$ of all the
infinitesimal increments $d\hat{y}$. $\langle
n|\rho_{\textrm{sys}}|n\rangle$ does not change as long as $\eta=0$,
and below we thus simply assume that $\xi_n(0)$ is the steady state
value. We note that this shift is insignificant if the changes of
the state of the system induced by the homodyne measurements take
place on a time scale that is much longer than $\kappa^{-1}$, which
is often the case, because the coupling strength $g$ of the
off-resonant interaction between the system and the cavity mode is
typically very small.

\subsection{Probing as a state preparation tool}

We next consider in more detail how the measurement of the phase
shift of the transmitted light can be used to prepare different
types of quantum mechanical states.

\subsubsection{Dicke states}

\begin{figure}
\begin{center}
\begin{tabular}{cc}
\multicolumn{1}{l}{\footnotesize\hspace{4mm}(a)}&
\multicolumn{1}{l}{\footnotesize\hspace{3mm}(d)}\\
\includegraphics*[viewport=4 2 392 298,width=0.48\columnwidth]{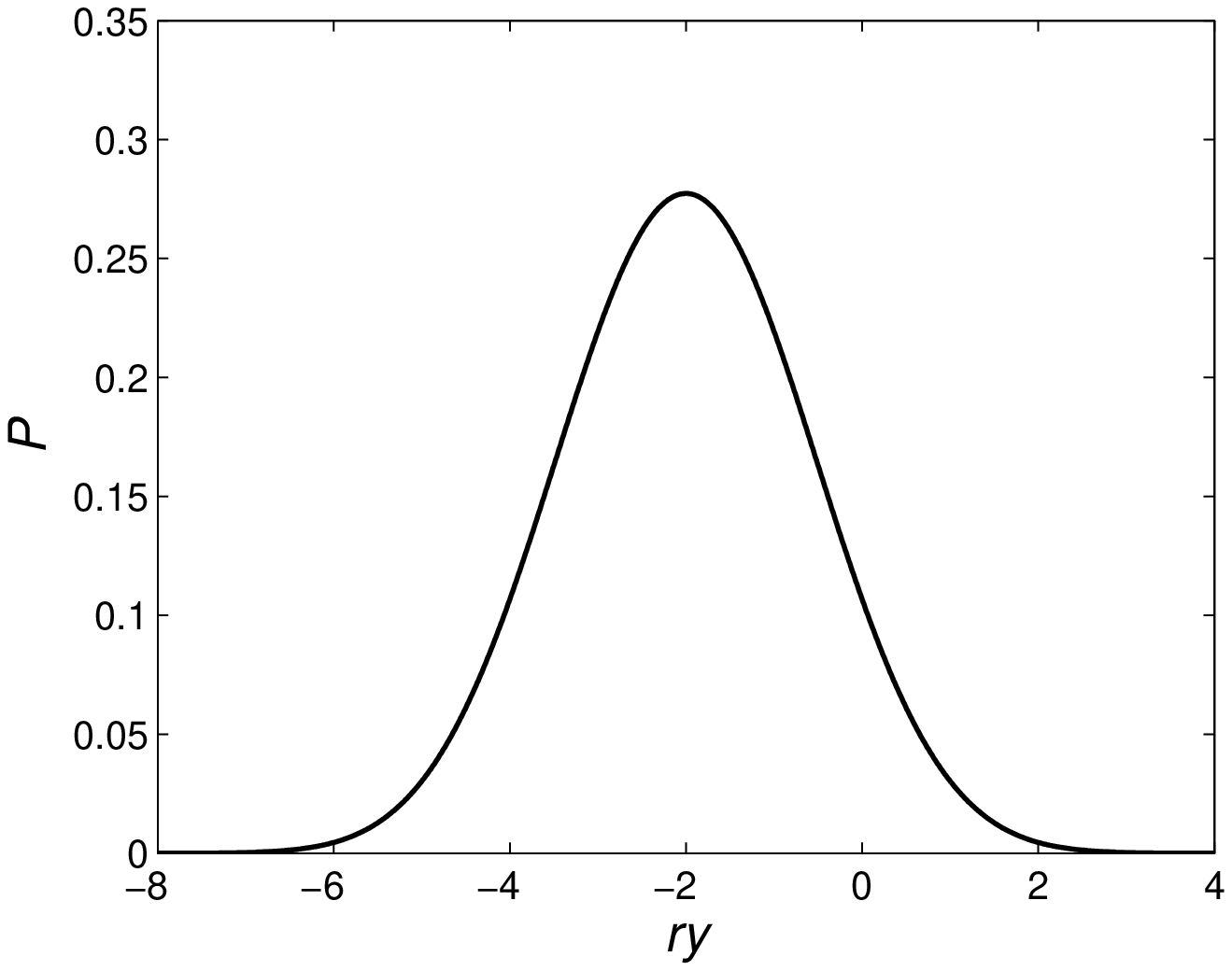}&
\includegraphics*[viewport=4 2 392 298,width=0.48\columnwidth]{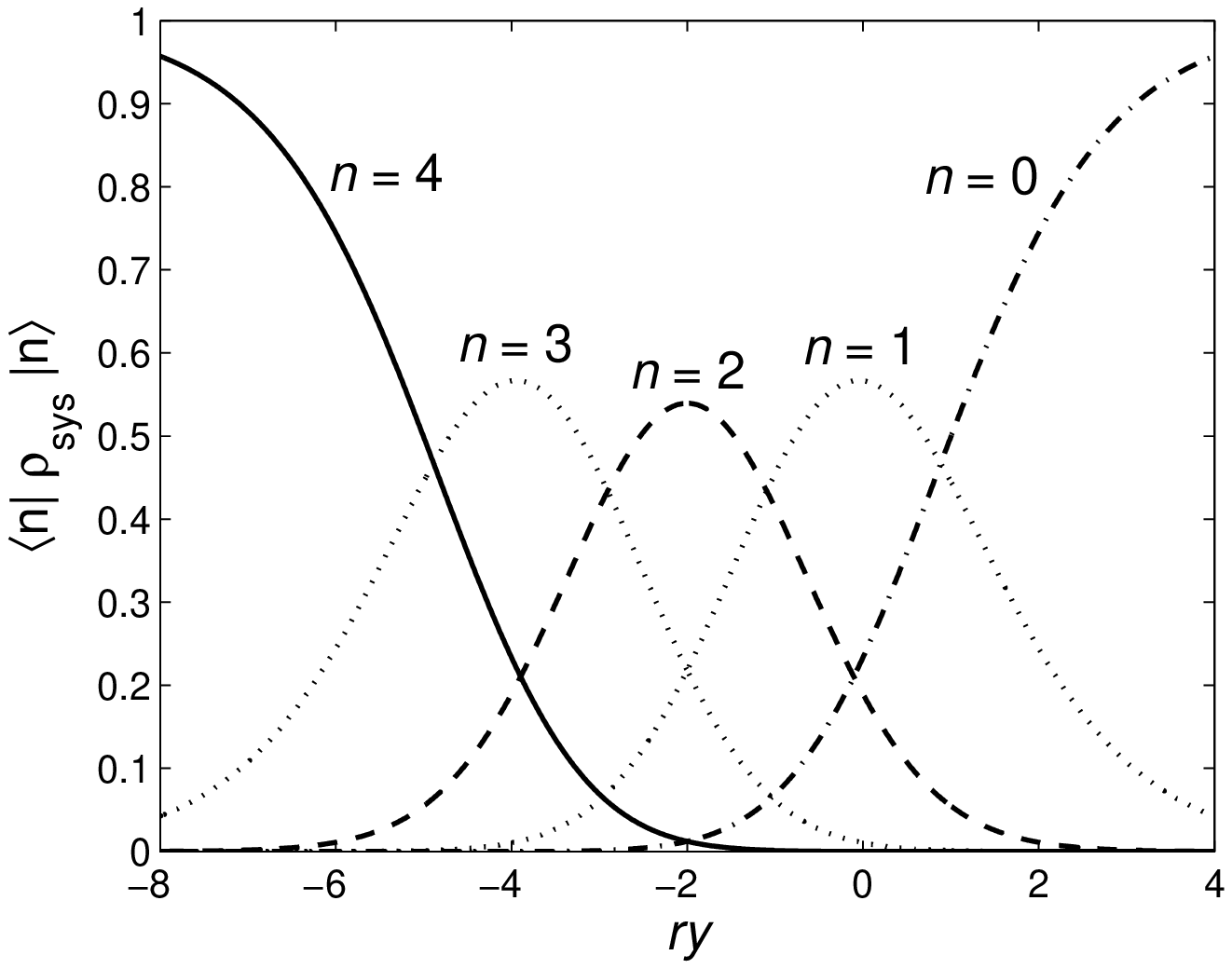}\\
\multicolumn{1}{l}{\footnotesize\hspace{4mm}(b)}&
\multicolumn{1}{l}{\footnotesize\hspace{3mm}(e)}\\
\includegraphics*[viewport=4 2 392 298,width=0.48\columnwidth]{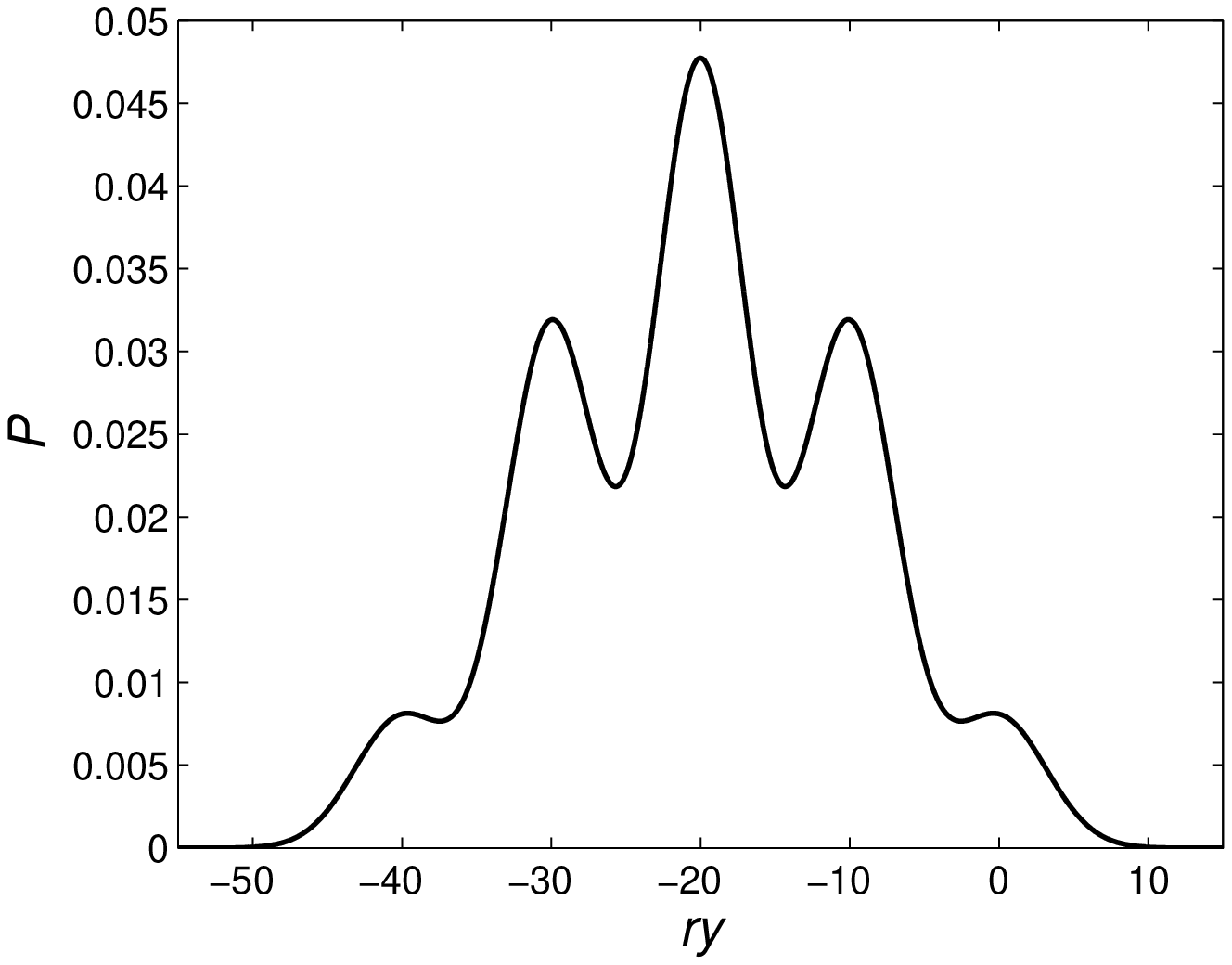}&
\includegraphics*[viewport=4 2 392 298,width=0.48\columnwidth]{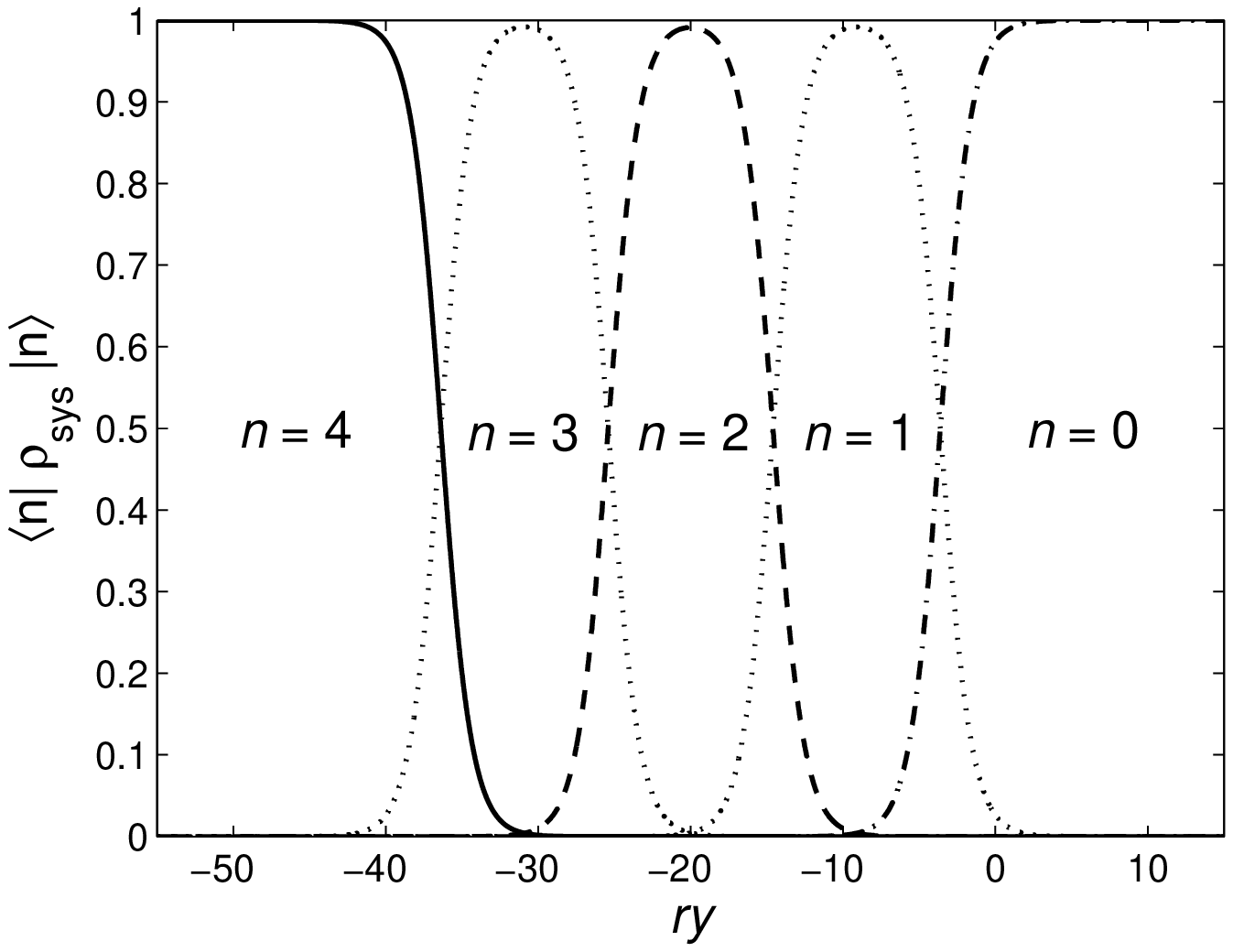}\\
\multicolumn{1}{l}{\footnotesize\hspace{4mm}(c)}&
\multicolumn{1}{l}{\footnotesize\hspace{3mm}(f)}\\
\includegraphics*[viewport=4 2 392 298,width=0.48\columnwidth]{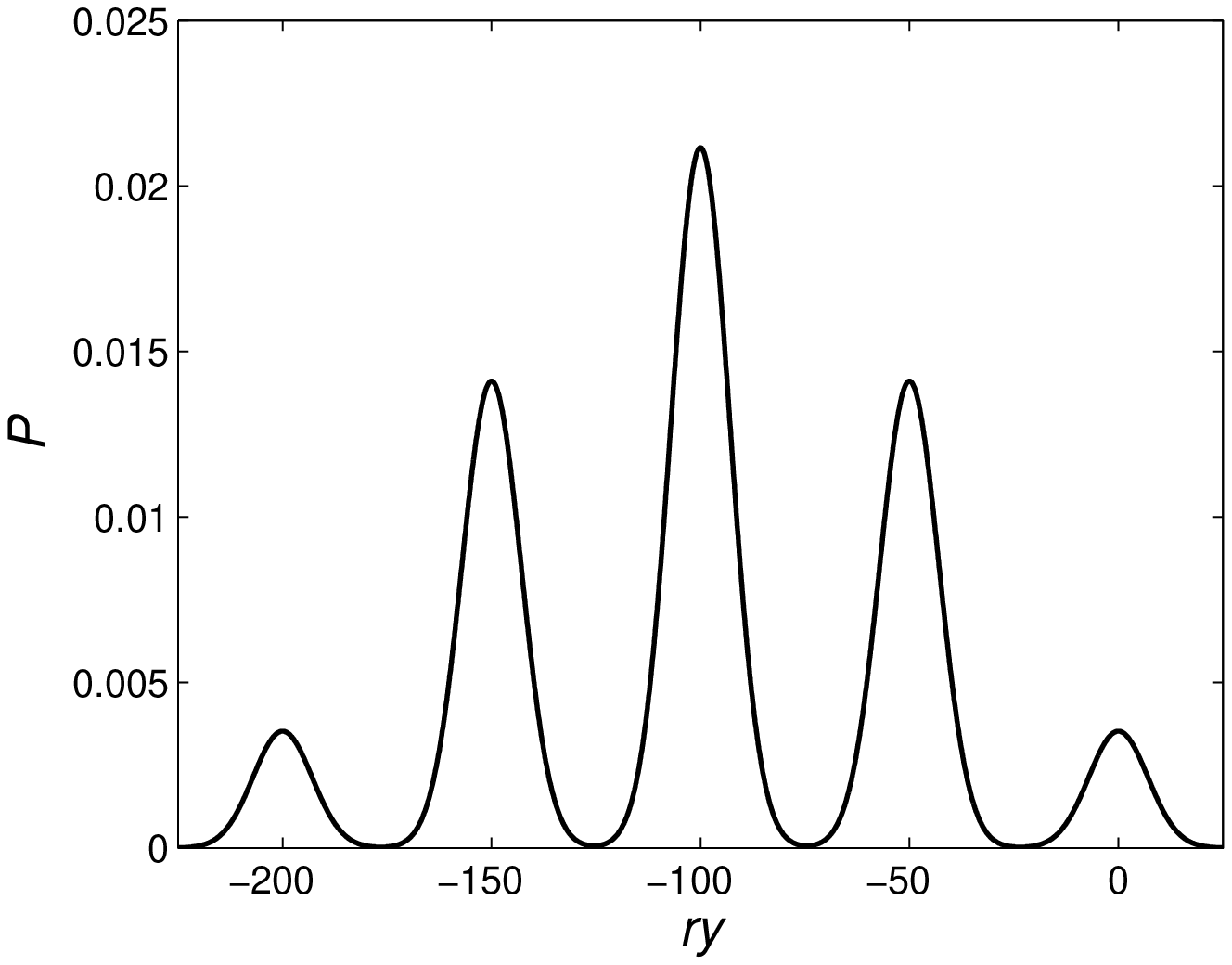}&
\includegraphics*[viewport=4 2 392 298,width=0.48\columnwidth]{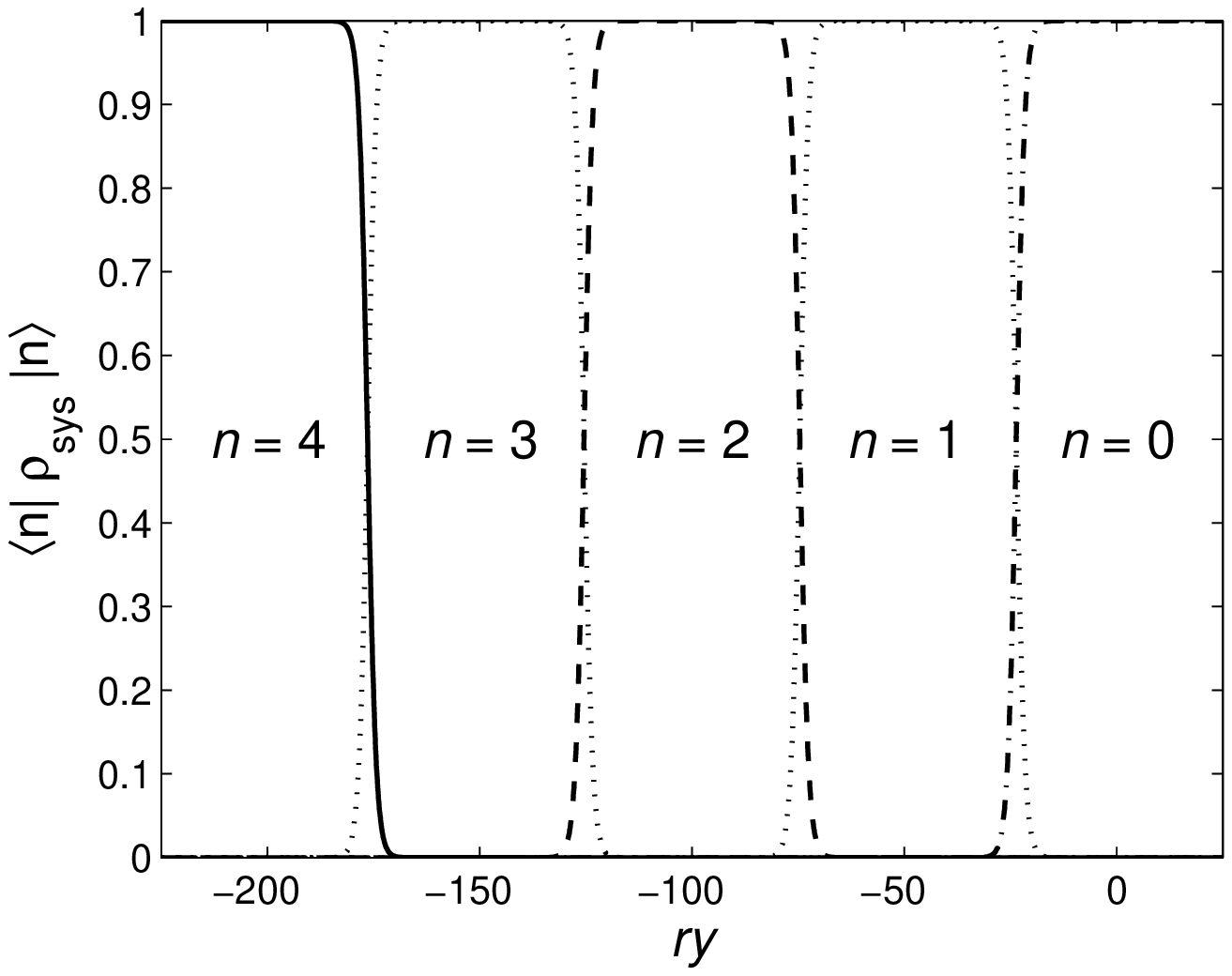}
\end{tabular}
\end{center}
\caption{(a)-(c): Probability density $P$ to measure the integrated
signal $y$ in the interval from $0$ to $t$ as a function of $y$ for
$r^2t=1$, $10$, and $50$, respectively. (d)-(f): $\langle
n|\rho_{sys}|n\rangle$ for $n=0,1,2,3,4$ as a function of $y$ for
the same values of $r^2t$. The values of $C_i(0)$ are those obtained
for the initial state
$|\psi_{\textrm{sys}}\rangle=((|f\rangle+|g\rangle)/\sqrt{2})^{\otimes
4}$ of the system, and it is assumed that $2g\ll\kappa$.}\label{N}
\end{figure}

To prepare a Dicke state we choose $\phi=0$ (corresponding to a
measurement of the $p$-quadrature of the field), since it follows
from Eq.\ \eqref{eq} and the definition of $r_n$ that those
measurements are most sensitive to the induced phase changes in the
limit $2g/\kappa\ll1$ and $\beta=\beta^*$. In this case
\begin{equation}
C_n(t)=C_n(0)\exp(-r_n\hat{y}-r_n^2t/2),
\end{equation}
with
\begin{equation}\label{rn}
r_n=\frac{8\beta gn\sqrt{\kappa_1\kappa_2\eta}}{\kappa^2+4g^2n^2},
\end{equation}
and, from Eq.\ \eqref{py}, the state for which $\hat{y}$ assumes the
value $y$ is obtained with probability density
\begin{equation}\label{PW}
P=\sum_{n=0}^N\frac{C_n(0)}{\sqrt{2\pi
t}}\exp\left(-\frac{(y+r_nt)^2}{2t}\right).
\end{equation}
$P$ and the state preparation fidelity $\langle
n|\rho_{\textrm{sys}}|n\rangle= C_n(t)/\sum_{m=0}^NC_m(t)$ are
plotted as functions of $y$ at different times for
$C_i(0)=N!/i!/(N-i)!/2^N$ and $N=4$ in Fig.\ \ref{N}, neglecting the
term $4g^2n^2$ in the denominator of Eq.\ \eqref{rn}, such that
$r_n=rn$, where $r$ is independent of $n$. Within this approximation
$P$ consists of a sum of five Gaussians separated by $rt$ and of
width $\sqrt{t}$, and the transition from overlapping Gaussians at
small $r^2t$ to well separated Gaussians at large $r^2t$ is apparent
in the figure. The plots of $\langle n|\rho_{\textrm{sys}}|n\rangle$
illustrate how the state of the system is gradually collapsed onto
an eigenstate of the operator $\hat{n}$, and it is clear that each
peak in $P$ corresponds to a specific value of $n$ if $r^2t$ is
large.

\subsubsection{Quantum superposition states}

It is also possible to use the setup to generate a superposition of
two Dicke states as, for instance, the maximally entangled state
$(|0\rangle+e^{i\theta}|N\rangle)/\sqrt{2}$. To do so we include an
additional phase shift inside the cavity such that the Hamiltonian
reads
\begin{equation}
H=\hbar g \hat{a}^\dag\hat{a}(\hat{n}-N/2).
\end{equation}
With this Hamiltonian the states $|n\rangle$ and $|N-n\rangle$ are
indistinguishable if the $x$-quadrature of the field is measured
($\phi=-\pi/2$ in Eq.\ \eqref{SME}), and thus, for the initial state
$|\psi_{\textrm{sys}}\rangle=((|f\rangle+|g\rangle)/\sqrt{2})^{\otimes
N}$, a superposition of $|0\rangle$ and $|N\rangle$ is generated
after sufficiently long measurement time with probability
$1/2^{N-1}$. To achieve a pure atomic state, it is required to work
under lossless conditions, i.e., under conditions where Eq.\
\eqref{SSE} is valid, and it is also important to avoid entanglement
between the cavity field and the atoms at the final time, since the
cavity field is to be traced out to obtain the atomic state. The
last requirement is easily fulfilled by turning off the probe field
at some point, letting the cavity field decay to the vacuum state
while keeping the detector turned on. We note that the protocol is
more demanding than the protocol to generate Dicke states due to the
requirement of lossless conditions and because the difference
between $r_n$ and $r_{n+1}$ is second order in $2g/\kappa$, which
leads to an increase in the required measurement time.

\subsection{Quantum Zeno effect and quantum jumps}

\begin{figure}
\begin{center}
\includegraphics*[viewport=12 2 388 300,width=0.95\columnwidth]{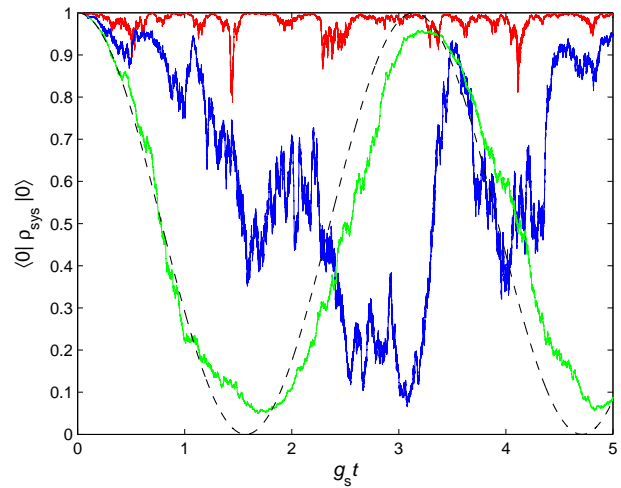}
\end{center}
\caption{$\langle0|\rho_{sys}|0\rangle$ as a function of scaled time
$g_\textrm{s}t$ for $N=1$, $\kappa_1=\kappa_2=0.5\textrm{ }\kappa$,
$\kappa_L=0$, $\eta=1$, $\phi=0$, $g=0.2\textrm{ }\kappa$,
$\beta=0.2\textrm{ }\sqrt{\kappa}$, $N_p=3$, and initial state
$\rho(0)=|0\rangle|vac\rangle\langle vac|\langle0|$, where
$|vac\rangle$ is the vacuum state of the cavity mode. For weak
driving with $g_s=0.001\textrm{ }\kappa$ (red curve) the continuous
measurements inhibit the coherent evolution between the states
$|0\rangle$ and $|1\rangle$, while for strong driving with
$g_s=0.05\textrm{ }\kappa$ (green curve) the time evolution of the
state almost follows the deterministic evolution obtained for $g=0$
(dashed curve). The blue curve ($g_s=0.005\textrm{ }\kappa$)
illustrates the behavior of the state of the system in the
intermediate regime. The same noise realization is used to compute
all curves.}\label{zeno}
\end{figure}

In the preceding subsections the state of the atoms was simply
detected by an indirect continuous measurement, but we now add the
complication of internal dynamics by allowing transitions between
adjacent $|n\rangle$ states. Specifically, we consider the
Hamiltonian
\begin{equation}\label{Hgs}
H=\hbar g\hat{a}^\dag\hat{a}\hat{n}+\hbar
g_s\sum_{i=1}^N(\hat{\sigma}_{i,+}+\hat{\sigma}_{i,-}),
\end{equation}
where $\sigma_{i,+}\equiv|f\rangle_{ii}\langle g|$ and
$\sigma_{i,-}\equiv|g\rangle_{ii}\langle f|$, and $i$ refers to atom
number $i$, which, depending on the level structure of the atoms,
may be realized by applying a microwave field propagating in a
direction perpendicular to the cavity axis. Since
\begin{equation}
\sum_{i=1}^N\hat{\sigma}_{i,+}|n\rangle=\sqrt{(n+1)(N-n)}|n+1\rangle,
\end{equation}
and
\begin{equation}
\sum_{i=1}^N\hat{\sigma}_{i,-}|n\rangle=\sqrt{n(N+1-n)}|n-1\rangle,
\end{equation}
we obtain the stochastic master equation for the system and the
cavity field by adding the terms
\begin{eqnarray}
&-ig_s\sqrt{n(N+1-n)}\rho_{n-1,m}&\label{gs1}\\
&-ig_s\sqrt{(n+1)(N-n)}\rho_{n+1,m}&\label{gs2}\\
&+ig_s\sqrt{m(N+1-m)}\rho_{n,m-1}&\label{gs3}\\
&+ig_s\sqrt{(m+1)(N-m)}\rho_{n,m+1}&\label{gs4}
\end{eqnarray}
to the right hand side of the master equation for the Hamiltonian in
Eq.\ \eqref{H}.

\begin{figure}
\begin{center}
\includegraphics*[viewport=12 2 388 300,width=0.95\columnwidth]{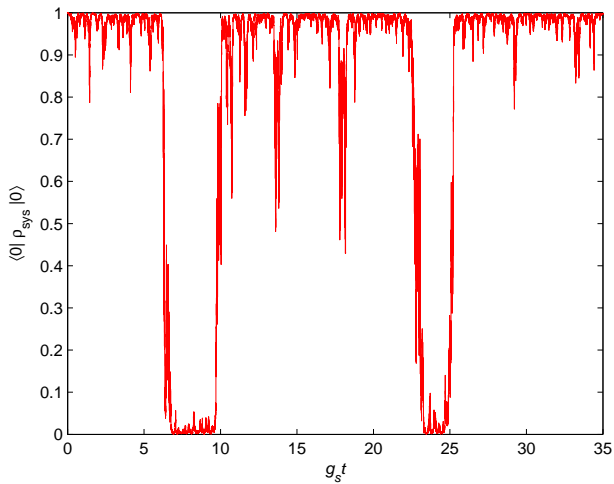}
\end{center}
\caption{Quantum jumps for a single atom in a cavity. The parameters
are the same as in Fig.\ \ref{zeno} ($g_s=0.001\textrm{
}\kappa$).}\label{qjump}
\end{figure}

The added system Hamiltonian drives a coherent evolution between the
different $|n\rangle$ states of the system, but if the measurements
are sufficiently strong, the continuous back action of the
measurements on the state of the system blocks the coherent
evolution. To demonstrate this effect, which is known as the Quantum
Zeno effect, we use the Milstein scheme (see Ref.\ \cite{kloeden})
to solve the \emph{nonlinear} stochastic master equation numerically
for a particular realization of the measurement outcomes. The
accuracy of the integration may be checked for $g_s=0$ by
application of the analytical results obtained in the previous
subsection. We use the Fock state basis for the cavity mode and
neglect terms in the density operator with more than $N_p$ photons
in the cavity, where $N_p\gg4\kappa_1|\beta|^2/\kappa^2$ (see Sec.\
\ref{VA}). The results are shown in Fig.\ \ref{zeno}. Since
$\langle1|\rho_{\textrm{sys}}|1\rangle$ is not exactly zero, there
is a small probability that the system switches to state
$|1\rangle$, where it is subsequently stabilized, and if the
dynamics is integrated for sufficiently long time, several such
quantum jumps between $|0\rangle$ and $|1\rangle$ will occur (Fig.\
\ref{qjump}).

We note that for general $N$ the states with $n=0$ or $n=N$ are
easier to stabilize by measurements, since the square root factors
in Eqs.\ \eqref{gs1}, \eqref{gs2}, \eqref{gs3}, and \eqref{gs4} are
smaller for $n$ close to $0$ or $N$ than they are for $n$ close to
$N/2$.

Finally, as a simple example of feedback, we note that it is
possible to generate a particular eigenstate of $\hat{n}$ by
increasing $g_s$ whenever $\langle n|\rho_{\textrm{sys}}|n\rangle$
becomes small and decreasing $g_s$ to zero whenever $\langle
n|\rho_{\textrm{sys}}|n\rangle$ becomes large.

\section{Conclusion}\label{VI}

In conclusion we have considered the influence of a specific
indirect continuous measurement on the state of an abstract system,
and we have derived the master equation determining the time
evolution of the state of the system conditioned on the measurement
outcomes. This equation provides a tool to analyze a wide variety of
systems and phenomena in detail, and we have applied it to
demonstrate the collapse of the state of a system onto an eigenstate
of the measurement operator and to demonstrate quantum jumps and
blocking of the internal dynamics of a system due to continuous
measurements.

Indirect measurements are important because they constitute a
special class of state manipulations that are very useful to prepare
systems in specific quantum mechanical states. The crucial point is
that the complete collapse of the state due to back action of a
measurement performed directly on the system is avoided by
performing instead the measurement on an auxiliary system that has
interacted with the system. We showed examples with simple atomic
level structures, but we emphasize that our analysis applies to
quite general quantum systems and, for example, to multi-level atoms
with shelving states leading to macroscopic quantum jumps in the
transmitted intensity \cite{metz,metz2}. A further step towards
achieving control on the quantum state of a system is to apply
feedback \cite{steck,geremia1,negretti} by subjecting the system to
disturbances that depend on the outcome of measurements. This
situation may also be handled by the formalism developed in the
present paper. The derived stochastic master equation is thus a
valuable tool to investigate further state preparation protocols.

\end{document}